# CHIMPS2: $^{13}$CO $J = 3 \rightarrow 2$ emission in the central molecular zone

S. M. King [ORCID],[1]* T. J. T. Moore,[1]* J. D. Henshaw,[1] S. N. Longmore,[1] D. J. Eden [ORCID],[2]* A. J. Rigby [ORCID],[3]
E. Rosolowsky [ORCID],[4] K. Tahani,[5] Y. Su,[6] A. Yiping,[6] X. Tang [ORCID],[7] S. Ragan [ORCID],[8] T. Liu,[9] Y.-J. Kuan[10,11]
and R. Rani[12]

[1]Astrophysics Research Institute, Liverpool John Moores University, IC2, Liverpool Science Park, 146 Brownlow Hill, Liverpool L3 5RF, UK
[2]Armagh Observatory and Planetarium, College Hill, Armagh BT61 9DB, UK
[3]School of Physics and Astronomy, University of Leeds, Leeds LS2 9JT, UK
[4]Department of Physics, University of Alberta, Edmonton, Alberta T6G 2E1, Canada
[5]Department of Physics & Astronomy, Kwantlen Polytechnic University, 12666 72nd Avenue, Surrey BC V3W 2M8, Canada
[6]Purple Mountain Observatory and Key Laboratory of Radio Astronomy, Chinese Academy of Sciences, Nanjing 210034, China
[7]Xinjiang Astronomical Observatory, 150 Science 1-Street, Urumqi, Xinjiang 830011, China
[8]School of Physics and Astronomy, Cardiff University, 5 The Parade, Newport Road, Cardiff CF24 3AA, UK
[9]Shanghai Astronomical Observatory, 80 Nandan Road, Xuhui District, Shanghai 200030, China
[10]Institute of Astronomy & Astrophysics, Academia Sinica. 11F Astronomy-Mathematics Building, AS/NTU No.1, Section 4, Roosevelt Road, Taipei 10617, Taiwan
[11]Department of Earth Sciences, National Taiwan Normal University, 88 Section 4, Ting-Chou Road, Taipei 11677, Taiwan, ROC
[12]Center of Astronomy and Gravitation, Department of Earth Sciences, National Taiwan Normal University, 88, Sec. 4, Ting-Chou Rd., Wenshan District, Taipei 116, Taiwan, ROC

## ABSTRACT

We present the initial data for the ($J = 3 \rightarrow 2$) transition of $^{13}$CO obtained from the central molecular zone (CMZ) of the Milky Way as part of the CO Heterodyne Inner Milky Way Plane Survey 2 (CHIMPS2). Covering $359° \leq l \leq 1°$ and $|b| \leq 0.5°$ with an angular resolution of 19 arcsec, velocity resolution of 1 km s$^{-1}$, and rms $\Delta T_A^* = 0.59$ K at these resolutions, our observations unveil the complex structure of the CMZ molecular gas in improved detail. Complemented by the $^{12}$CO CHIMPS2 data, we estimate a median optical depth of $\tau_{13} = 0.087$. The preliminary analysis yields a median $^{13}$CO column-density range equal to $N(^{13}\text{CO}) = 2$–$5 \times 10^{18}$ cm$^{-2}$, median H$_2$ column density equal to $N(\text{H}_2) = 4 \times 10^{22}$ cm$^{-2}$ to $1 \times 10^{23}$ cm$^{-2}$. We derive $N(\text{H}_2)$-based total mass estimates of $M(\text{H}_2) = 2$–$6 \times 10^7$ M$_\odot$, in agreement with previous studies. We analyse the relationship between the integrated intensity of $^{13}$CO and the surface density of compact sources identified by *Herschel* Hi-GAL, and find that younger Hi-GAL sources detected at 500 μm but not at 70 μm follow the dense gas of the CMZ more closely than those that are bright at 70 μm. The latter, actively star-forming sources appear to be more associated with material in the foreground spiral arms.

**Key words:** stars: formation – ISM: molecules – ISM: structure – Galaxy: centre – submillimetre: ISM.

## 1 INTRODUCTION

Galaxies are home to a wide variety of environments, all of which could conceivably influence the process of star formation. Our own Milky Way Galaxy contains large-scale variations in the Galactic environment from the outer reaches of the spiral arms to the inner plane and central molecular zone (CMZ). Star formation from molecular gas is one of the key driving forces behind Galactic evolution and morphology, yet current understanding of how the efficiency of this process [the star formation efficiency (SFE)] is regulated on local and Galactic spatial scales is little understood.

The fractional abundances of molecular gas within the Galaxy decrease rapidly from ∼100 per cent within 1 kpc of the centre to only a few per cent at Galactocentric distances ∼10 kpc (Sofue & Nakanishi 2016). The inner 500 parsec of the Galaxy, known as the CMZ, contains some of the most extreme environments in which star formation can occur. When compared to the Galactic disc, the average density, temperature, and velocity dispersion of this region are found to be several orders of magnitude larger (Henshaw et al. 2023). However, within the CMZ there exists a large quantity of dense molecular gas, with mass ∼ $10^7$ M$_\odot$, and mean density ∼ $10^4$ cm$^{-3}$ (Morris & Serabyn 1996; Ferrière, Gillard & Jean 2007). Given our current understanding of the star formation process, the CMZ should be producing a significant number of stars; yet observations have shown that the region produces stars with an efficiency at least a factor of 10 times lower than the spiral arms (Longmore et al. 2013).

Recently, advanced array detectors have enabled large-scale, efficient surveys of the Galaxy, allowing us to study the impact on the star formation efficiency of regions such as the spiral arms (Urquhart et al. [2018]). However, unravelling the intricacies of star formation on smaller spatial scales is more complicated. Empirical relations such as Kennicutt–Schmidt (Kennicutt [1998]), which correlates the star formation rate (SFR) with the surface density of molecular gas in galaxies on large spatial scales, break down on the smaller scales where the enclosed sampling of the initial mass function (IMF) is small. The CMZ should be fully sampled in this regard but fails to be fully represented by current empirical relations. (Kruijssen & Longmore [2014]).

To better examine the physical properties and processes that give rise to star formation in the CMZ, it is necessary to conduct large-scale observational surveys of molecular clouds across multiple isotopologues and transitions, across a diverse range of star-forming environments, such as the $^{13}$CO/C$^{18}$O Heterodyne Inner Milky-Way Plane Survey (CHIMPS; Rigby et al. [2016]), conducted using the Heterodyne Array Receiver Programme (HARP) detector (Buckle et al. [2009]) via the James Clark Maxwell Telescope between $28° \leq l \leq 46°$ and $|b| \leq 0.5°$ in the rotational transition $J = 3 \rightarrow 2$ to better identify the dominant mechanism(s) that drive the process.

The follow-up survey, CHIMPS2, builds on the observations made by CHIMPS, again focusing on the $J = 3 \rightarrow 2$ transition of $^{13}$CO and C$^{18}$O, extending the inner-plane coverage and adding the CMZ and the outer plane, including $^{12}$CO in the latter regions. Other surveys covering CO emission within the CMZ include the Structure, Excitation, and Dynamics of the Inner Galactic Interstellar Medium Survey (SEDIGISM; Schuller et al. [2017]), observed using the APEX telescope in $^{13}$CO and C$^{18}$O $J = 2 \rightarrow 1$, at a resolution of 30 arcsec, longitude range $-60°$ to $+18°$, and latitudes $|b| < 0.5°$, and the Large-Scale CO Survey of the Galactic Centre conducted with the Nobeyama 45-m telescope in $^{12}$CO and $^{13}$CO $J = 1 \rightarrow 0$ (Oka et al. [1998]). The latter covers longitudes $-1.5°$ to $+3.4°$ and latitudes within $|b| < 0.6°$.

Additionally, surveys covering other molecules include; the H$_2$O Southern Galactic Plane Survey (HOPS; Walsh et al. [2011]; Longmore et al. [2013]) observed using the Mopra Radio Telescope covering Galactic longitudes $290° < l < 360°$ and $0° < l < 30°$ at 19.5–27.5 GHz, and the Combined Array for Research in Millimeter-wave Astronomy (CARMA; Pound & Yusef-Zadeh [2018]) 3-mm survey which mapped the CMZ between $-0.2 \leq l \leq 0.5$ and $|b| \leq 0.2$. The frequency range of HOPS covers spectral-line emission from H$_2$O masers, as well as metastable transitions of multiple molecules such as ammonia (NH$_3$), cyanoacetylene (HC$_3$N), methanol (CH$_3$OH), and radio recombination lines. While CARMA covers transitions of several molecules, SiO ($J = 2 \rightarrow 1$), HCO$^+$ ($J = 1 \rightarrow 0$), HCN ($J = 1 \rightarrow 0$), N$_2$H$^+$ ($J = 1 \rightarrow 0$), and CS ($J = 2 \rightarrow 1$).

In this paper, we present $^{13}$CO $J = 3 \rightarrow 2$ data from the January 2023 version of the CMZ portion of the CHIMPS2 survey. We present first-order estimates of the physical properties of molecular gas traced by these data and assess the implications of the results in the context of SFE. The structure of the paper is as follows. Section 2 describes the data, Section 3 provides an analysis of the $^{13}$CO emission, and Section 4 provides a summary of the results. The corresponding $^{12}$CO data can be seen in Eden et al. [2020].

## 2 OBSERVATIONS AND DATA REDUCTION

The CMZ portion of the CHIMPS2 survey employed a systematic observation strategy, as described in Eden et al. [2020], covering approximately two square degrees constructed from several individual tiles. Each tile is $21 \times 21$ arcmin in dimension, positioned 20 arcmin apart, with each tile overlapping to facilitate calibration adjustments and correct for edge effects; this pattern is easily visible in Fig. 1.

The observations included here cover Galactic longitudes $359° \leq l \leq 1°$, latitudes $|b| \leq 0.5$, have a spatial resolution of 16 arcsec, spectral resolution of 1 km s$^{-1}$ over velocities of $|V_{LSR}| \leq 300$ km s$^{-1}$ and an rms in $T_A^*$ of 0.59 K on 6-arcsec pixels. The 1-GHz bandwidth mode of the ACSIS spectrometer used with HARP was employed to cover the extended velocity range of approximately 550 km s$^{-1}$ in the CMZ (Dame, Hartmann & Thaddeus [2001]).

As this bandwidth is larger than that used for the rest of the CHIMPS2 survey and precluded simultaneous tuning of ACSIS to the C$^{18}$O line, only $^{13}$CO was observed in this portion of the survey. Additional C$^{18}$O $J = 3 \rightarrow 2$ data were taken later as follow-up observations towards regions of bright $^{13}$CO and will be the subject of a future publication.

### 2.1 Data reduction and masking

#### 2.1.1 Data reduction

Reduction of both the $^{12}$CO and $^{13}$CO data was carried out using the automated pipeline ORAC-DR (Jenness & Economou [2015]), which had been improved to meet the specific requirements of the survey. A detailed description of the data reduction process can be found in Eden et al. [2020].

To create comprehensive spectral cubes covering the entire observation area, the individual tile cubes were combined using the WCSMOSAIC function of the Starlink applications software suite, from its 2021A release (Currie et al. [2014]). Although the data cover a velocity range of $V_{LSR} = \pm 300$ km s$^{-1}$, no emission was detected in velocity channels outside $|V_{LSR}| = 250$ km s$^{-1}$; therefore, to create the mosaic seen in Fig. 1, the data cube was collapsed in the range $V_{LSR} = \pm 250$ km s$^{-1}$ so as not to integrate unnecessary noise, resulting in a single integrated-intensity map (position, position) of the CMZ.

#### 2.1.2 Masking

The signal-to-noise ratio (SNR) in each pixel of the integrated-intensity mosaic for both $^{12}$CO and $^{13}$CO data sets was calculated using the Starlink MAKESNR function. This function creates a SNR map from the input mosaic by dividing the data by the square root of its variance array. A binary mask was then created using the THRESH function, which applies a specified threshold to an input SNR map. The threshold was set to SNR = 5; values below this threshold were set to 0 and values at or above this threshold were set to 1. The binary mask was applied to the original integrated intensity mosaic by multiplying the two, effectively retaining the values of pixels with SNR $\geq 5$ and setting those $< 5$ at zero. Finally, the $^{12}$CO mosaic was aligned with the $^{13}$CO map using WCSALIGN to remove areas that were not observed in $^{13}$CO.

The intensities presented in the integrated emission images in Fig. 1 are on the $T_A^*$ scale. For conversion to the main-beam temperature scale, $T_{mb}$, the main-beam efficiency $\eta_{mb}$ was assigned a value of 0.64 and 0.78 for $^{12}$CO and $^{13}$CO, respectively (Buckle et al. [2009]).

## 3 RESULTS AND ANALYSIS

In this section, we present the initial results of the $^{13}$CO $J = 3 \rightarrow 2$ CMZ portion of the CHIMPS2 survey. These data provide an initial look at the molecular clouds within the CMZ, traced to a higher density, and are combined with data from the $^{12}CO$ CMZ portion of the CHIMPS2 survey (Eden et al. 2020) for analysis and comparison.

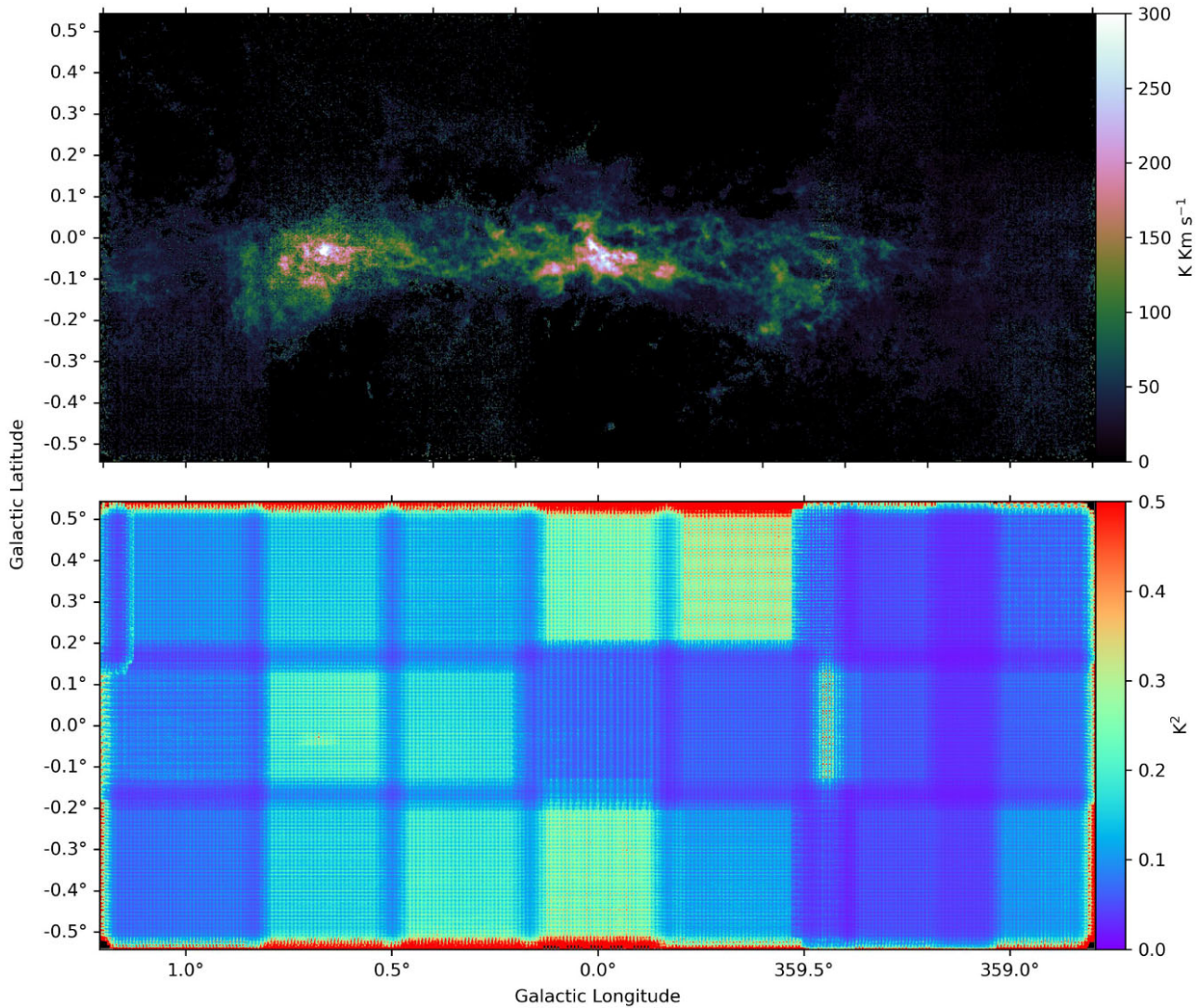

**Figure 1.** Top: The integrated emission mosaic of the CMZ traced by the $J = 3 \rightarrow 2$ transition of $^{13}CO$, each spectrum was integrated over velocity channels between $|V_{LSR}| = 250$ kms$^{-1}$. Bottom: The variance map of the $^{13}CO$ $J = 3 \rightarrow 2$ data shown in the top panel.

### 3.1 Integrated emission map

Panel (a) of Fig. 1 shows the integrated intensity mosaic of $^{13}CO$ $J = 3 \rightarrow 2$ emission in the CMZ between $l = 359°$ and $l = 1°$, and $|b| \leq 0°.5$, while panel (b) shows the variance mosaic of the $^{13}CO$ integrated intensity map in panel (a), representative of the relative noise levels associated with each tile in the mosaic. The $^{13}CO$ map shows many similarities to the $^{12}CO$ map in Eden et al. (2020) in terms of overall morphology; however, the $^{13}CO$ data exhibit several notable differences. As $^{13}CO$ traces denser regions of molecular clouds, the majority of the emission from the more diffuse regions of the CMZ is not present. Extended filamentary structures throughout the region are much clearer in $^{13}CO$, due to the reduced optical depth, demonstrating the hierarchical structure of molecular clouds.

#### 3.1.1 Constraining the $^{12}CO$ to $^{13}CO$ abundance ratio

Fig. 2 shows the ratio of integrated $^{12}CO$ to $^{13}CO$ intensities in the CMZ. The lower values, characterized by blue features, indicate regions with relatively high $^{13}CO$ intensity and show the deeper hierarchical structure of the CMZ within the context of the brighter $^{12}CO$ surrounding it. The intensity ratios of $^{12}CO$ to $^{13}CO$ are directly related to the abundance ratios of $^{12}C$ to $^{13}C$, modified by optical depth. Therefore, it is possible to constrain the abundance ratio of the two molecules using the observed intensity ratio $R$:

$$R = \frac{T_{12}}{T_{13}} \approx \frac{1 - e^{-\tau_{12}}}{1 - e^{-\tau_{13}}}. \tag{1}$$

If both species are optically thick, the brightness temperature ratio $R$ approaches 1. If both are optically thin, then $R$ approaches $\tau_{12}/\tau_{13}$, which equals the $^{12}CO/^{13}CO$ abundance ratio. Therefore, since $\tau_{13} \ll \tau_{12}$, the largest observed intensity ratio will occur where $\tau_{12}$ is lowest, making $R$ closest to the $^{12}CO/^{13}CO$ abundance ratio, the upper limit of this ratio is shown in Fig. 3.

The upper limit was established at $R \sim 24$ using the 95th percentile of the observed values, to delineate the upper end of the observed values, filtering out potential noise outliers, and ensuring that our analysis is conservative and representative of the CMZ. This gives a lower limit to the abundance ratio of $^{12}CO$ to $^{13}CO$, as the line ratio may still be suppressed by significant $^{12}CO$ optical depth. The value is consistent with previous estimates of the abundance ratio in the CMZ, which vary between 20 and 25 (Penzias 1981; Guesten, Henkel & Batrla 1985; Langer & Penzias 1990; Wilson & Matteucci 1992; Langer & Penzias 1993; Riquelme et al. 2010; Belloche et al. 2013; Humire et al. 2020).

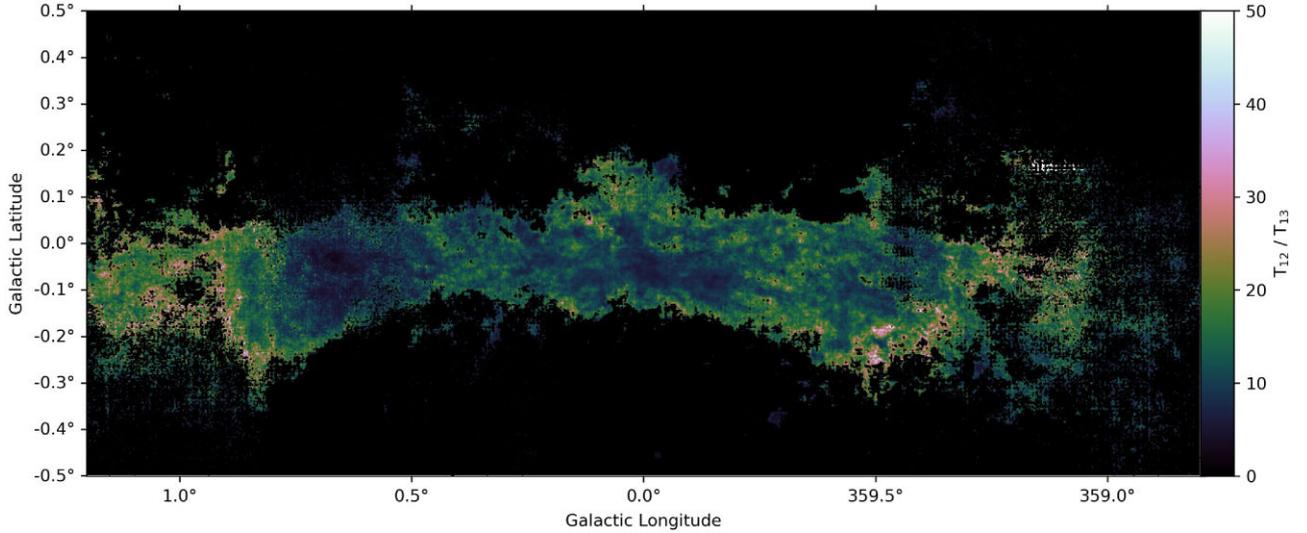

**Figure 2.** Showing the ratio of $^{12}$CO to $^{13}$CO integrated-emission maps. The low-intensity regions are those with high concentrations of $^{13}$CO and are associated with a colder, denser molecular gas.

### 3.2 $^{13}$CO optical depth

The optical depth of the velocity-averaged $^{13}$CO3 → 2 line ($\tau_{13}$) was determined from the integrated intensity ratio as described by the following equations, in a similar method as that shown in equation (1) resulting from the solution of the equation of radiative transfer:

$$\frac{T^*_{A,12}}{T^*_{A,13}} = \frac{\eta_{12}}{\eta_{13}} \frac{\phi_{12}}{\phi_{13}} \frac{T_{ex,12}}{T_{ex,13}} \frac{1 - e^{-\tau_{12}}}{1 - e^{-\tau_{13}}}, \tag{2}$$

where $\eta_{12}$, $\eta_{13}$ are the relative main-beam efficiencies for $^{12}$CO and $^{13}$CO, $\phi_{12}$ and $\phi_{13}$ are the relative beam-filling factors for each species, $\tau_{12}$ is the optical depth in the $^{12}$CO (3 → 2) line; finally, $T_{ex,12}$ and $T_{ex,13}$ are the excitation temperatures for $^{12}$CO and $^{13}$CO, respectively.

Since the main rotational constants and hence the transition energies of the two species differ by only about 4.5 per cent, it adds a negligible systematic error to assume a common excitation temperature for both. Additionally, given that $^{12}$CO and $^{13}$CO are spatially coexistent and trace similar regions within the molecular clouds, it is reasonable to assume that their emissions fill the beam similarly, so that $\eta_{12} \approx \eta_{13}$, at least to first order. So, by assuming equal beam-filling factors and excitation temperatures for both species, and considering $\tau_{12}$ to be optically thick, equation (2) simplifies to

$$\frac{T^*_{A,12}}{T^*_{A,13}} \simeq \frac{\eta_{12}}{\eta_{13}} \frac{1}{1 - e^{-\tau_{13}}}. \tag{3}$$

Consequently, the optical depth can then be expressed as

$$\tau_{13} \simeq -\ln\left(1 - \frac{T^*_{A,13}}{T^*_{A,12}} \frac{\eta_{12}}{\eta_{13}}\right). \tag{4}$$

To obtain the pixel-to-pixel optical-depth map in Fig. 4, equation (4) was applied to the ratio of $T^*_{A,12}$ to $T^*_{A,13}$ for each value in Fig. 2. The median value in Fig. 4 is $\tau_{13} = 0.087$ and a per-pixel random error of $\pm 5 \times 10^{-4}$ was determined by standard error propagation. Systematic errors arising from the assumptions above are likely to be several times larger than this. This median value suggests optically thin $^{13}$CO emission throughout most of the CMZ. The optical depths here are averages over the line widths at each point due to the use of the integrated intensity, and underestimate peak $\tau_{13}$. When $^{13}$CO is optically thin, equation (2) becomes

$$\frac{T_{mb,12}}{T_{mb,13}} = R \simeq \frac{1 - e^{-\tau_{12}}}{\tau_{13}}. \tag{5}$$

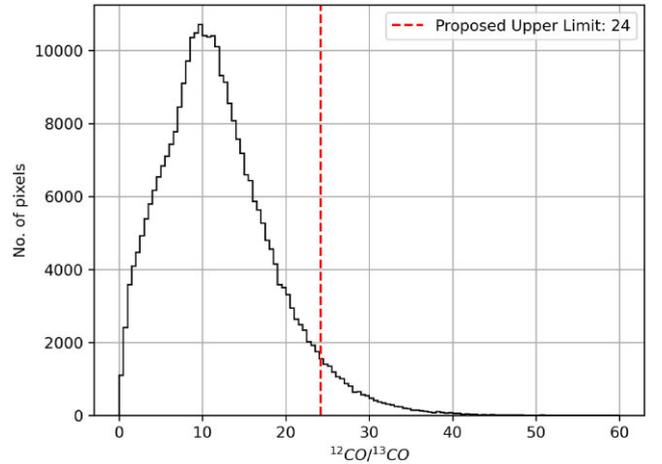

**Figure 3.** The distribution of intensities in $^{12}$CO to $^{13}$CO ratio map of the CMZ, highlighting the 95th percentile threshold, which establishes the lower limit of the abundance ratio for $^{12}$CO to $^{13}$CO.

Where $T_{mb,12}$ and $T_{mb,13}$ are the corrected antenna temperatures converted to main-beam temperature for simplicity and assuming equal excitation temperatures and beam filling factors. Therefore, where $\tau_{12} > 1$.

$$\tau_{13} > \frac{1 - e^{-1}}{R}, \tag{6}$$

which sets a lower bound on $\tau_{13}$, based on the intensity ratio $R$, above which $^{12}$CO will be optically thick.

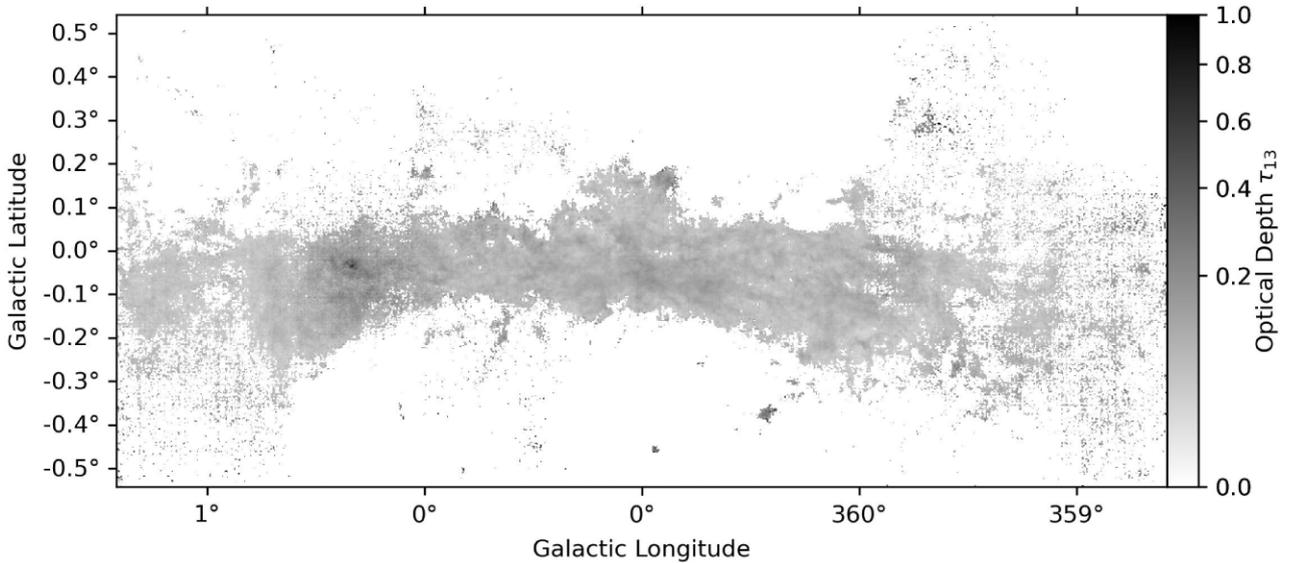

**Figure 4.** The $^{13}$CO $J = 3-2$ optical depth in the CMZ, calculated using the ratio of intensity values of $^{12}$CO/$^{13}$CO. The CMZ clearly exhibits mainly optically thin emission, with only a small number of regions appearing to become optically thick, in regions such as Sgr B2. The maximum non-noise associated optical depth determined was $\tau_{13} = 0.99$.

Given that the maximum reliable observed value of $R$ is $\sim$24 (Section 3.1.1), $\tau_{12}$ is > 1 where $\tau_{13}$ > 0.0263. The minimum $\tau_{13}$ value determined in this study was found to be approximately 0.035. Therefore, it is likely that $^{12}$CO is mostly optically thick even though the $\tau_{12} \gg 1$ assumption of equation (3) leads to slight overestimates of $\tau_{13}$ where this is not the case.

### 3.2.1 Optical depth in the Sgr B2 region

Fig. 4 shows that most regions in the CMZ exhibit relatively low $^{13}$CO $J = 3-2$ optical depths. Sgr B2 accounts for a considerable portion of the star formation activity and molecular mass in the CMZ. The expectation is that such a dense and active region would exhibit higher optical depths due to the large column densities of molecular gas. However, the optical depth of $^{13}$CO remains relatively low over much of the Sgr B2 region, only becoming optically thick towards the centre. A likely explanation for this is averaging over large line widths due to the use of integrated intensities in equation (4) Figs B1 and B2 show the distribution of gas velocities across the CMZ. In particular, the gas velocities in Sgr B2 are shown to reach up to $\sim 100 \, \mathrm{km \, s^{-1}}$ (Fig. B2). Typical line widths in the Sgr B2 region are $\sim 20 \, \mathrm{km \, s^{-1}}$ much broader than in the Galactic spiral arms (Ao, Yang & Sunada 2004; Rigby et al. 2019). The optical depths in the line peaks are likely to be several times higher.

### 3.3 $^{13}$CO column density

For optically thin emission, it was possible to determine a first-order approximation of the column density for this region, as the solution to the equation of radiative transfer also describes the column density of the $^{13}$CO, $J = 3 \to 2$ line. Under the assumption of local thermodynamic equilibrium (LTE), the optical depth is proportional to the column density of the gas along the line of sight, for the high-temperature approximation of the partition function $Z = kT/hB$ and, assuming a constant excitation temperature, the column density

for $^{13}$CO can be expressed as

$$N(^{13}\mathrm{CO}) = \frac{8\pi}{7c^2} \frac{kT_\mathrm{ex}}{hB} \frac{\nu_{32}^2}{A_{32}} e^{\frac{6hB}{kT_\mathrm{ex}}} \left(1 - e^{-\frac{h\nu_{32}}{kT_\mathrm{ex}}}\right)^{-1} \int \tau_{13} \, \mathrm{d}\nu. \quad (7)$$

Here, $c$ is the speed of light, $B$ represents the rotation constant in frequency units, $B = 55.10 \, \mathrm{GHz}$, $k$ and $h$ are the Boltzmann and Planck constants, respectively, $T_\mathrm{ex}$ is the excitation temperature for the $J = 3 \to 2$ transition of $^{13}$CO, and $\nu_{32}$, $A_{32}$, and $\tau_{13}$ are the transition frequency, Einstein coefficient, and $^{13}$CO optical depth calculated using equation (4), respectively.

To estimate the column density range for $^{13}$CO, we assume an excitation temperature range of $T_\mathrm{ex} = 20$–$50 \, \mathrm{K}$. This assumption is based on the findings of previous studies (Morris et al. 1983; Martin 2006; Nagai et al. 2007), which provide well-established temperature ranges for similar molecular environments. Adopting these values allows for consistency with the literature and ensures that our results are consistent with those of existing research.

Using this temperature range, the column densities range from $(2 \pm 0.01) \times 10^{18} \, \mathrm{cm^{-2}}$ to $(5 \pm 0.03) \times 10^{18} \, \mathrm{cm^{-2}}$. The small uncertainties are the random errors arising from the measurement uncertainties, via the calculated optical depths, which have of $\Delta\tau_{13} = 5 \times 10^{-4}$. The uncertainty in $T_\mathrm{ex}$ is clearly much larger, as will other systematic factors such as calibration accuracy, abundance ratios, filling factors, and assumptions inherent in the LTE model.

### 3.4 $H_2$ mass calculation

By considering the commonly accepted abundance ratio of $^{12}$CO to $H_2$, $\alpha_{12} = 10^{-4}$ for the Milky Way (Frerking, Langer & Wilson 1982; Solomon et al. 1987; Dame et al. 2001), and the adopted lower-limit abundance ratio (see Section 3.1.1), we were able to constrain the $^{13}$CO to $H_2$ abundance ratio to $\alpha_{13} \leq 4 \times 10^{-5}$, 40 times larger than the $^{13}$CO to $H_2$ ratio given for the solar neighbourhood of $1 \times 10^{-6}$ (Frerking et al. 1982). From this, the conversion to $H_2$ column density was simple, such that

$$N(H_2) = \frac{N(^{13}\mathrm{CO})}{\alpha_{13}}. \quad (8)$$

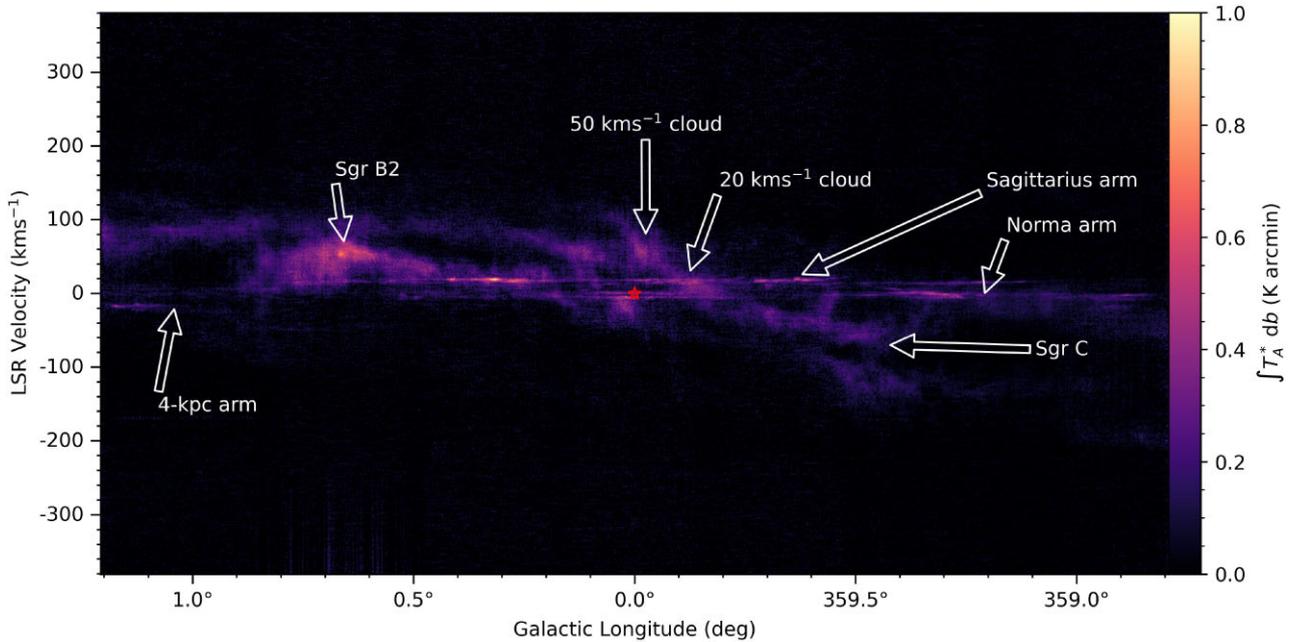

**Figure 5.** Showing the longitude–velocity map of $^{13}CO\,J = 3 \rightarrow 2$ in the CMZ, the intensity being integrated over latitude from data complete as of Oct 2023. The foreground spiral arms appear as linear features with low velocity dispersion. The presence of emission from the arms in this data set is likely to skew calculation of the physical properties determined for the CMZ molecular gas. The red star indicates the position of Sgr A*.

The median $H_2$ column density range was found to be between $N(H_2) = (4 \pm 0.02) \times 10^{22}$ and $(1 \pm 0.003) \times 10^{23}\,cm^{-2}$. The highest values of $N(H_2)$ reached between $5 \times 10^{24}$ and $1 \times 10^{25}\,cm^{-2}$ for $T_{ex} = 20\,K$ and $50\,K$, respectively.

For a lower temperature limit of $T_{ex} = 20\,K$, our calculations yield a total molecular-hydrogen mass of $M(H_2) = (2 \pm 0.4) \times 10^7\,M_\odot$. Assuming an upper temperature limit of $T_{ex} = 50\,K$ provides a mass estimate of $M(H_2) = (6 \pm 1) \times 10^7\,M_\odot$. The associated uncertainties on each of the mass estimates again represent the propagation of random errors in our measurements. The lower mass in this range is consistent with estimates reported in previous surveys of approximately $2 \times 10^7\,M_\odot$ (Bally et al. 1988; Nagai et al. 2007), and $3 \times 10^7\,M_\odot$ returned by Bally et al. (1987) and Dahmen et al. (1998) for a region larger than that of this study. This result is consistent with the estimated range reported by Morris & Serabyn (1996) of $5–10 \times 10^7\,M_\odot$, suggesting that the true $T_{ex}$ for the gas traced by the $^{13}CO\,J = 3 \rightarrow 2$ line lies at the lower end of the temperature range.

However, assuming a constant $\alpha_{13}$ and $T_{ex}$ is likely to overestimate the column density in regions with lower $^{13}CO$ to $H_2$ abundances, while underestimating it in regions with a higher abundance ratio. The same can be said for the assumption of constant $T_{ex}$, as it is unlikely that the temperature is uniform across the CMZ. It would therefore be more accurate to derive $T_{ex}$ using assumptions under LTE and the solution to the equation of radiative transfer, which would return a more suitable $T_{ex}$ than applying a temperature range. This will be the topic of a subsequent paper.

### 3.5 Kinematic structure

#### 3.5.1 High-velocity features

Fig. 5 shows the position–velocity $(P - V_{LSR})$ distribution of the $^{13}CO\,J = 3 \rightarrow 2$ line integrated over the entire latitude range. Inspection of the high velocity dispersion features towards $l \approx 0°$ shows several distinct features such as Sgr C, along with the 20- and

50-km s$^{-1}$ clouds. Perhaps the most noticeable feature in Fig. 5 is Sgr B2, a complex and massive molecular cloud featuring high gas densities, and velocities between $0 \leq V_{LSR} \leq -100\,km\,s^{-1}$. Sgr B2 is one of the prominent features within the CMZ, it is one of the most massive molecular clouds in the Galaxy, and is known for its rich, intricate structure, and elevated star formation rate (Scoville, Solomon & Penzias 1975; Möller et al. 2023).

#### 3.5.2 Foreground features

Some of the most apparent features in Fig. 5 are the Far Sagittarius, 4-kpc, and Outer Spiral arms that can be seen as bright and narrow lines of emission with a velocity distribution $|V_{LSR}| = 20\,km\,s^{-1}$ throughout the figure. These features were seen in emission and absorption in the $^{12}CO$ data (Eden et al. 2020), but in emission here, since $^{13}CO$ is not subject to the self-absorption commonly seen in the $^{12}CO$ data. These features are emission from at least two spiral arms, probably the Sagittarius arm, which can be seen spanning discontinuously across the image at $V_{LSR} \approx 10\,km\,s^{-1}$. A small section of the 4-kpc arm is visible at $V_{LSR} \approx -20\,km\,s^{-1}$ towards $l \approx 1°.2$. Finally, the outer arm is present ($V_{LSR} \approx -20\,km\,s^{-1}$) at $l \approx 359°.5$. Each of these spiral arms are a contamination source for the $^{13}CO$ CMZ data, and calculation of the physical properties in Section 3 includes them, which skews the CMZ results.

### 3.6 Hi-GAL source distribution in the CMZ

To investigate the distribution of Hi-GAL sources in the CMZ, we selected two distinct samples from the Hi-GAL compact-source catalogue. The '70-μm-bright' subset contains sources detected at 70 μm as well as the other Hi-GAL bands. The '70-μm-dark' set includes sources detected in all bands except for 70 μm. The two subsets represent sources that are actively star forming and those that are younger, potentially star forming or not star forming.

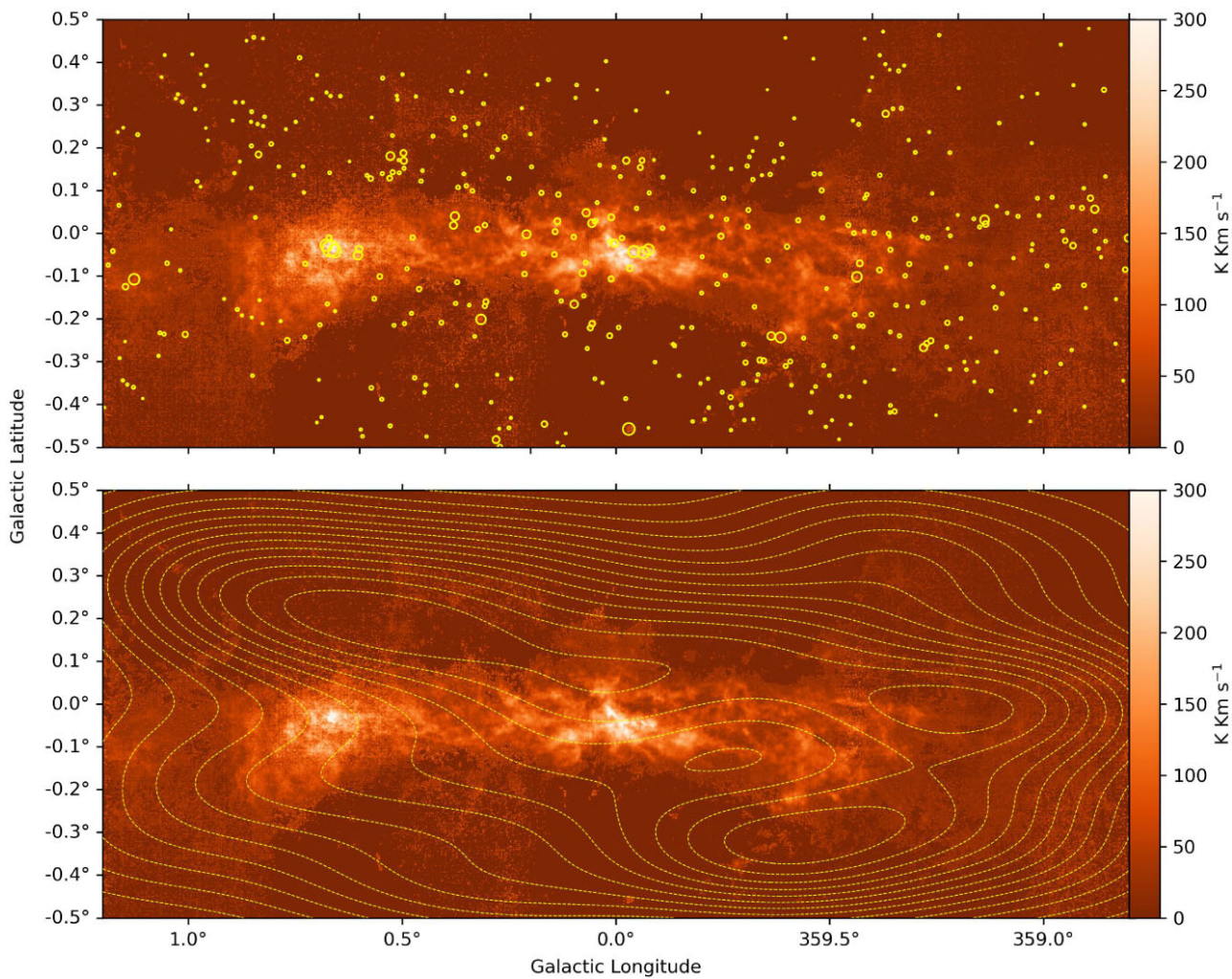

**Figure 6.** The integrated $^{13}$CO $J = 3 - 2$ intensity from Fig. 1, overlaid with (top panel) the positions of 70-$\mu$m-bright compact sources (circles) detected by Hi-GAL (Molinari et al. 2016a) and (bottom panel) corresponding relative source surface-density contours. The size of each point in the top panel is proportional to the square root of the 70-$\mu$m source flux density.

### 3.6.1 Hi-GAL source density and distribution

The upper panel of Fig. 6 displays the distribution of compact sources identified by Hi-GAL (Molinari et al. 2016a) that are detected at 70 $\mu$m and considered to be active star-forming sources (Molinari et al. 2016b). The source positions are overlaid on to the integrated $^{13}$CO intensity map (Fig. 1). The lower panel of Fig. 6 presents contours of the relative surface density of 70-$\mu$m-bright sources, again overlaid on the integrated $^{13}$CO intensity distribution. Notably, there is a decrease in the concentration of sources towards lower latitudes, with higher source densities found above and below the centre line. This suggests that fewer protostellar sources are associated with the CMZ molecular gas itself, with higher concentrations possibly being associated with foreground spiral-arm clouds.

Fig. 7 shows the distribution of the 70-$\mu$m-dark sample and reveals a different trend in the locations of these sources, which tend to follow the general shape of the $^{13}$CO CMZ emission, suggesting a closer relationship between the two. These objects, potentially represent an earlier stage of star formation such as pre-stellar cores, compared to the 70-$\mu$m-bright catalogue. The distribution of these sources correlates better with the structural features of the CMZ, possibly tracing the dense gas associated with the dust ridge, but is fairly uniform at higher latitudes.

### 3.6.2 Two-dimensional Kolmogorov–Smirnov testing

To test the relationship between the $^{13}$CO integrated intensity and the Hi-GAL source density for each catalogue, a modified Kolmogorov–Smirnov (KS) test was used to test the relationship in two dimensions (Peacock 1983), as opposed to the traditional KS testing which is one-dimensional. This was done to preserve the spatial information in the relationship. The $^{13}$CO integrated intensity distribution was compared to the 70-$\mu$m-bright and dark source densities separately. The significance level was set at 0.003. This threshold means that if the returned $p$-value is less than 0.003, the null hypothesis (that the two distributions are from the same underlying continuous distribution) is rejected with 99.7 per cent confidence.

For the 70-$\mu$m-bright sources, the KS test yielded a statistic of 0.56 with a $p$-value of $\sim$0, suggesting a significant statistical divergence between the distributions of $^{13}$CO integrated emission and Hi-GAL source density. This result confirms the obvious difference in the nature of the data sets; however, considering that star formation predominantly occurs within molecular clouds, a degree of spatial correlation between the $^{13}$CO integrated emission and the locations of protostellar objects could be expected, despite the fact that they do not originate from the same underlying population.

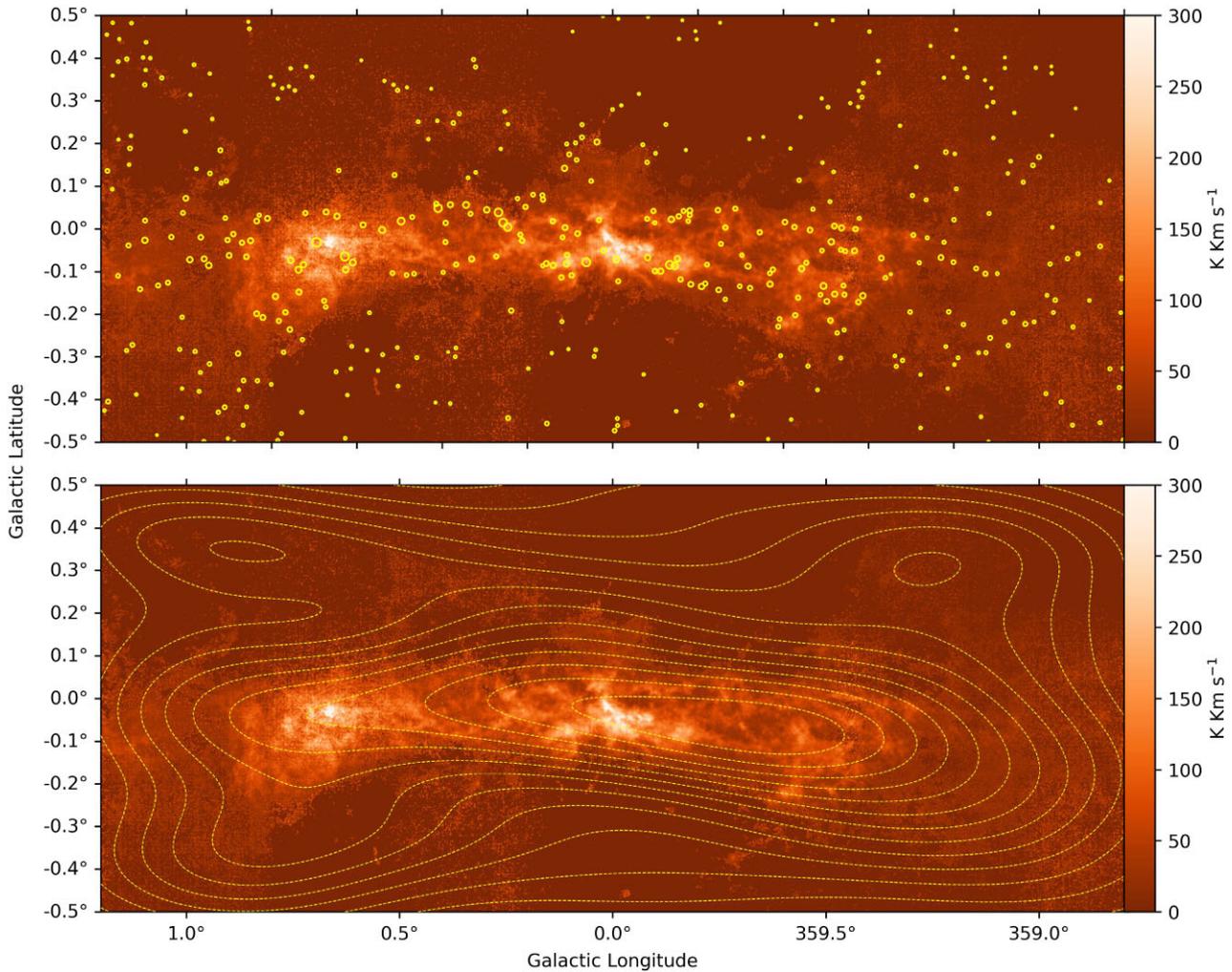

**Figure 7.** As for Fig. 6, but for Hi-GAL compact sources not detected at 70 μm.

Furthermore, Spearman correlation analysis of the $^{13}$CO integrated emission with the Hi-GAL source surface density at corresponding pixels showed moderate but significant negative correlation. In this case, $\rho_{bright} = -0.35$ with associated $p$-value $= 0.0003$. This suggests that higher densities of 70-μm-bright sources tend to occur in areas with lower concentrations of $^{13}$CO. Therefore, the statistical results support the visual impression of the relative spacing of the contours in Fig. 6, and suggest that the potentially protostellar 70-μm-bright sources are not preferentially associated with the dense gas of the CMZ, implying that they are mainly in the foreground.

In contrast, analysis of the 70-μm-dark sources produces a somewhat different result. The KS statistic returned was 0.55, again with $p \sim 0$, indicating a degree of divergence similar to that for the 70-μm-bright sources. However, Spearman's correlation for this data set ($\rho_{dark} = 0.58$, $p = 1 \times 10^{-9}$) points to a significant positive relationship, suggesting a closer spatial association between the 70-μm-dark sources and regions with significant $^{13}$CO integrated emission and that the younger sources tend to be more associated with the $^{13}$CO emission from the CMZ itself.

The differences in the source populations are reflected in the correlation analysis, where the 70-μm-bright sources show a moderate negative correlation with $^{13}$CO emission, suggesting they are often in less dense regions or in the foreground. In contrast, the 70-μm-dark sources exhibit a moderate positive correlation, indicating a strong association with the dense gas in the CMZ. The distinct distributions of 70-μm-bright and 70-μm-dark sources towards the CMZ with respect to the molecular gas suggest variations in the star formation stages across different latitudes. However, considering the potential contamination from the foreground spiral arms, as indicated in Fig. 5, a non-negligible number of these sources might not be inherent to the CMZ.

The presence of spiral arms complicates the interpretation of the relationships observed in our data. Although we have identified statistically significant correlations between the surface densities of both 70-μm-bright and 70-μm-dark sources and the $^{13}$CO integrated intensity, the extent to which these correlations are influenced by the spiral-arm structures requires closer scrutiny. To assess the implications of these correlations for the SFE in the CMZ, a more detailed analysis that accounts for the contributions from spiral arms is essential. Future work will therefore focus on reevaluating the data with the spiral arms removed.

We also conducted a direct comparison between the distributions of 70-μm-bright and 70-μm-dark sources. The results showed a KS statistic of 0.64 with a *p*-value of $\sim 0$, indicating a significant statistical difference between the two data sets. This firmly rejects the null hypothesis of a common distribution, corroborating the notion that these source populations are distinct. Furthermore, the correlation analysis between the 70-μm-bright and 70-μm-dark sources yielded a Spearman coefficient of $-0.18$, with a *p*-value of 0.074. This result suggests that there is no significant correlation between these data sets, further supporting the conclusion that the two data sets are likely to represent different populations.

## 4 SUMMARY

In this paper, we have presented initial findings from the $^{13}$CO $J = 3 \rightarrow 2$ observations of the CMZ made as part of the CHIMPS2 survey. The data used in this study are available to download from the CANFAR archive. These observations offer a detailed view of molecular clouds within the CMZ, tracing higher densities at higher resolution and lower optical depth than previous CO surveys. The integrated intensity maps of $^{13}$CO in the CMZ shown in Fig. 1 reveal intricate physical structures of the colder, denser gas.

By combining these $^{13}$CO measurements with the $^{12}$CO data also from the CHIMPS2 survey (Eden et al. 2020), we obtain estimates of the optical depth, $\tau_{13}$ distribution in the region, leading to a first-order estimate of the $^{13}$CO column density, $N(^{13}\text{CO}) = (2 \pm 0.01) \times 10^{18}$ cm$^{-2}$ to $N(^{13}\text{CO}) = (5 \pm 0.03) \times 10^{18}$ cm$^{-2}$. This corresponds to H$_2$ column density estimates of $N(\text{H}_2) = (4 \pm 0.02) \times 10^{22}$ cm$^{-2}$ to $(1 \pm 0.003) \times 10^{23}$ cm$^{-2}$. We present total H$_2$ mass estimates, $M(\text{H}_2) = (2 \pm 0.4) \times 10^7$ M$_\odot$ and $M(\text{H}_2) = (6 \pm 1) \times 10^7$ M$_\odot$ for $T_{ex} = 20$ K and $T_{ex} = 50$ K, respectively. This range is consistent with previous studies but likely to be an overestimate due to the assumption of constant $\alpha_{13}$ and $T_{ex}$ for this study. Analysis of the kinematic structure uncovers high-velocity-dispersion features, such as the well-known Sagittarius A and B complexes, and emission from foreground spiral arms, probably including the Sagittarius, Outer, and 4-kpc arms.

We find that compact *Herschel* Hi-GAL sources that are detected at 70 μm tend to avoid the molecular gas of the CMZ itself in favour of material that is likely to be in the foreground spiral arms. Since such objects are considered active sites of star formation, this is consistent with observations of relatively low star formation efficiency in the dense gas of the CMZ (Longmore et al. 2013). Conversely, we find that the distribution of the Hi-GAL sources without 70 μm emission tend to follow the dense-gas distribution more closely.

These high-resolution observations of $^{13}$CO within the CMZ represent a valuable resource for future CMZ studies, especially when combined with forthcoming C$^{18}$O CHIMPS2 data. Integrating these data sets with other molecular gas surveys at lower transitions, such as SEDIGISM (Schuller et al. 2021), along with the Hi-GAL continuum data (Molinari et al. 2016b), will provide gas temperature estimates and improved column density results. This work will be the subject of a series of future papers. Additional future work will include analysis of the separated signals from the spiral arms and CMZ molecular gas to better assess the differences between the two components and allow analysis of the turbulence parameters of dense molecular gas in the CMZ.

## ACKNOWLEDGEMENTS

We thank the anonymous referee for their comments and commitment to improving the clarity of this article.

The *James Clerk Maxwell Telescope* is operated by the East Asian Observatory on behalf of The National Astronomical Observatory of Japan; Academia Sinica Institute of Astronomy and Astrophysics; the Korea Astronomy and Space Science Institute; the Operation, Maintenance and Upgrading Fund for Astronomical Telescopes and Facility Instruments, budgeted from the Ministry of Finance (MOF) of China and administrated by the Chinese Academy of Sciences (CAS). Additional funding support is provided by the Science and Technology Facilities Council of the United Kingdom and participating universities in the United Kingdom and Canada. The Starlink software (Currie et al. 2014) is currently supported by the East Asian Observatory. This research has made use of NASA's Astrophysics Data System.

## DATA AVAILABILITY

The reduced CHIMPS2 $^{13}$CO CMZ data are available to download from the CANFAR archive. The data are available as mosaics, roughly $2° \times 1°$ in size, as well as individual observations. Integrated $l - b$ maps and $l - V_{LSR}$ maps, displayed in Figs 1 and 5 for the whole CMZ are provided, as well as the $l - V_{LSR}$ maps for the individual cubes. The data are presented in FITS format.

## APPENDIX A: VOXEL DISTRIBUTION

The distribution of the voxel values in the data cube can be seen in the top panel of Fig. A1. The distribution can be modelled by a Gaussian curve (red line) with associated mean $T_A^* = 0.016$ K and standard deviation $\sigma = 0.23$ K. The top panel inset shows the distribution on a log-$y$ scale to illustrate how the distribution departs from a standard Gaussian shape in the positive and negative wings due to real emission and non-Gaussian noise in the positive wing, and non-uniform and non-Gaussian noise in the negative wings. The bottom panel of Fig. A1 displays the values the square root of values in the variance mosaic shown in the bottom panel of Fig. 1, thus giving the standard deviation. The distribution peaks at $\approx 0.26$ K comparable to the value obtained from the standard deviation of the fit in panel (a).

## APPENDIX B: ADDITIONAL FIGURES

A breakdown of the $^{13}$CO integrated-emission map can be seen in Figs B1 and B2, where the data have been separated into velocity channels of 50 km s$^{-1}$. Fig. B1 shows several structures within the $-150$- to 100-km s$^{-1}$ velocity channel which are high-velocity $^{13}$CO surrounding Sgr C, which itself sits within the $-100$- to $-50$-km s$^{-1}$ channel. Additionally, the three cores of the Sgr A complex are most prominent in the $-50$- to 0-km s$^{-1}$ channel, which are the bright regions located at $l \sim 0°$.

Most of the emission is concentrated within the velocity range $0 - 100$ km s$^{-1}$ visible in Fig. B2, which contains features such as the 20-km s$^{-1}$ cloud, the 50-km s$^{-1}$ cloud and the 'brick', which are characterized by intense CO emission with sharp boundaries. Several prominent features are shown, including the shell-like structure and emission 'hole' in the 0–50-km s$^{-1}$ channel that was first identified by Bally et al. (1988), while 50–100 km s$^{-1}$ contains the high-velocity gas closer to Sgr B2. The 100–150 km s$^{-1}$ panel of Fig. B2 interestingly shows $^{13}$CO emission from a compact region in the centre of the Galaxy; the intense emission here is likely to be high-velocity molecular gas surrounding Sgr A*.

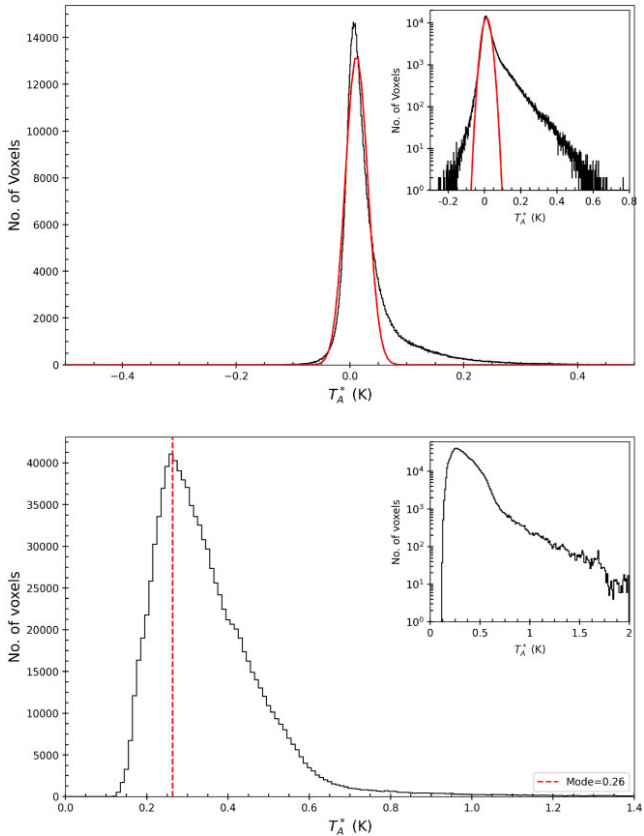

**Figure A1.** Top: A histogram of the voxel values in the original $^{13}$CO $J = 3$–2 data cube used to create the integrated intensity map shown in panel (a) of Fig. 1. The inset panel shows the same data on a log-$y$ scale, to highlight the wings of the distribution. Bottom: As for the top panel, but showing the standard deviation of variance values in panel (b) of Fig. 1. The peak (red dashed line) is $T_A^* = 0.26$ K. The inset shows the same distribution on a log-$y$ scale.

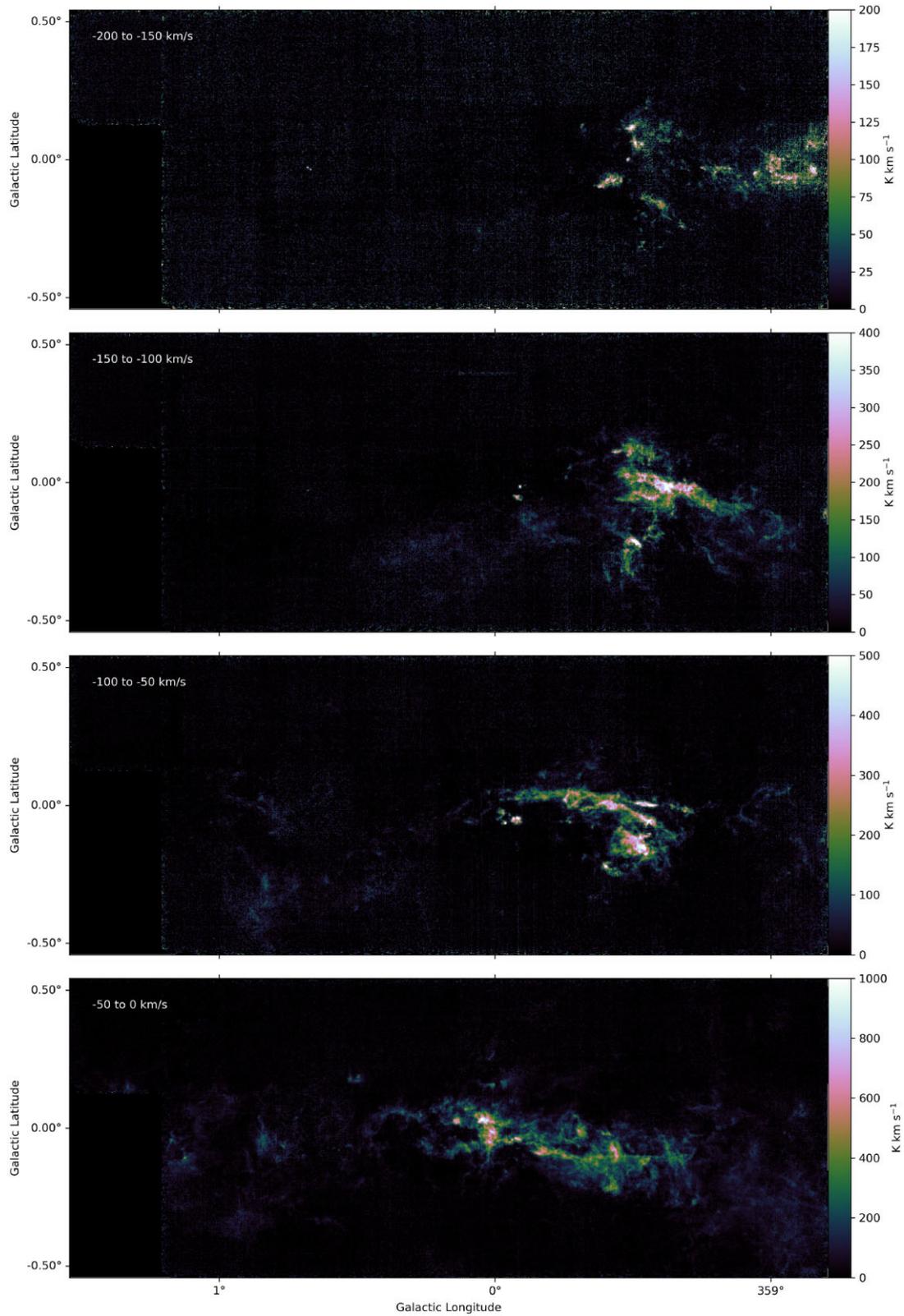

**Figure B1.** The $^{13}$CO $J = 3 \rightarrow 2$ integrated emission separated into 50 km s$^{-1}$ channels. The top map is $-250$ to $-200$ km s$^{-1}$; the second panel is $-200$ to $-150$ km s$^{-1}$; the third is $-150$ to $-100$ km s$^{-1}$; and the fourth is $-100$ to $-50$ km s$^{-1}$.

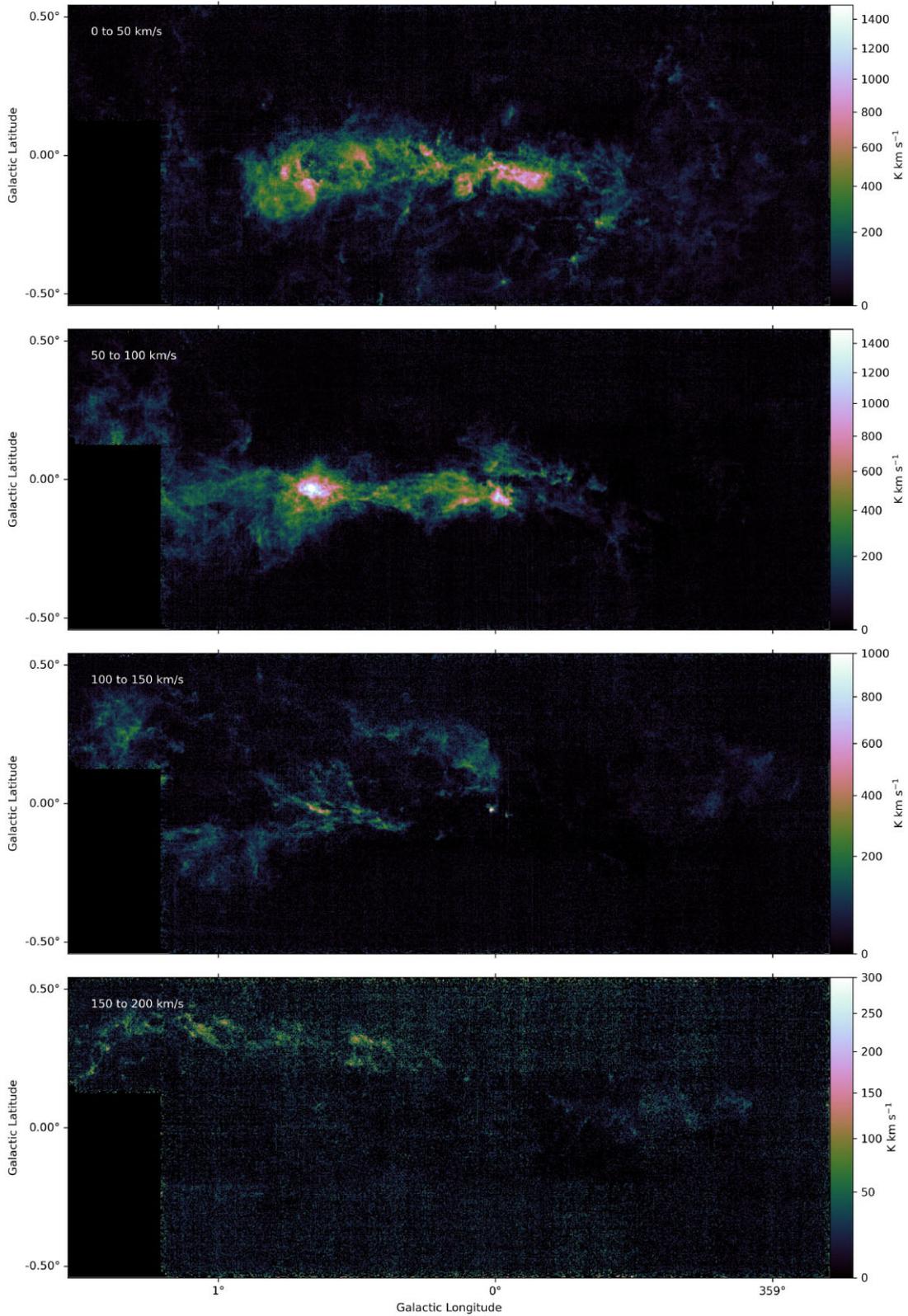

**Figure B2.** A continuation of the 50 km s$^{-1}$ channels of integrated emission. These are −50 to 0 km s$^{-1}$; 0 to 50 km s$^{-1}$; 50 to 100 km s$^{-1}$; and 100 to 150 km s$^{-1}$, respectively.

This paper has been typeset from a T$_{E}$X/L$^{A}$T$_{E}$X file prepared by the author.